\documentclass[twocolumn,secnumarabic,amssymb,amsmath, nobibnotes, aps, pra]{revtex4-1}

\usepackage{graphicx}% Include figure files
\graphicspath{{images/}{pictures/}{figures/}}
\usepackage{amsmath}
\begin{document}

\title{Self-consistent Purcell factor and spontaneous topological transition in hyperbolic metamaterials} %Title of paper

\author{Sergey Krasikov$^1$}
\author{Ivan V.~Iorsh$^1$} 
%\email{i.iorsh@phoi.itmo.ru}
\affiliation{$^1$ ITMO University, St.~Petersburg 197101, Russia}

\date{\today}

\begin{abstract}
In this work we develop a self-consistent approach for calculation of the Purcell factor and Lamb shift in highly dispersive hyperbolic metamaterial accounting for the effective dipole frequency shift.
Also we theoretically predict the possibility of spontaneous topological transition, which occurs not due to the external change of the system parameters but only due to the Lamb shift.

\end{abstract}

\maketitle
\section{INTRODUCTION}
The Purcell effect is the ratio of the spontaneous emission rate of a dipole placed inside the media in comparison to the rate in vacuum~\cite{purcell1946spontaneous}. This ratio can be tuned by engineering the environment of the source and the possibility of controllable change of radiative lifetime has been already demonstrated in a variety of different systems \citep{tanaka2010multifold,shubina2010plasmon,jun2008nonresonant,rao2007single,%
yao2009ultrahigh,noginov2010controlling}.
One of the most promising systems for this purpose are hyperbolic metamaterials --- highly anisotropic uniaxial media with dielectric permittivity tensor principal components of different signs~\cite{smith2003electromagnetic}. This defines the hyperboid shape of the isofrequency contours for the extra-ordinary waves, which leads to a number of unique properties and as consequence to the  variety of applications~\cite{poddubny2013hyperbolic}. %
In particular, hyperbolic isofrequency surface leads to a broadband singularity in photonic density of states so the Purcell factor is huge~\cite{jacob2010engineering,jacob2012broadband}. It has been later shown that there are different mechanisms to regularize the density of states singularity: finite emitter size~\cite{poddubny2011spontaneous}, finite size of the metamaterial unit cell~\cite{iorsh2012spontaneous}, and nonlocal response~\cite{yan2012hyperbolic}. However experimental studies of the Purcell effect in hyperbolic metamaterials~\cite{krishnamoorthy2012topological,noginov2010controlling,ni2011loss,kim2012improving} still demonstrate sufficient discrepancies with the theoretical predictions. In this Letter we show that there is another source of the distinction between the theoretical predictions and measurements results originating of the strong frequency dispersion of the metamaterial effective parameters. Namely we show that in the case of small emitters and strong frequency dispersion the conventional approach based on the evaluating of the Green's function at the position of the dipole and at the bare emitter frequency provides sufficiently inexact results and  we introduce the rigorous self-consistent method which  should be used instead. Moreover, we predict the effect of the Lamb-shift induced topological transition in metal-dielectric metamaterials, where the interaction of the emitter with the back reflected emitted electromagnetic fields leads to the effective pulling of the emitter inside outside of  the hyperbolic regime.
\section{MODEL}
\begin{figure}
\begin{center}
\includegraphics[width=0.9\linewidth]{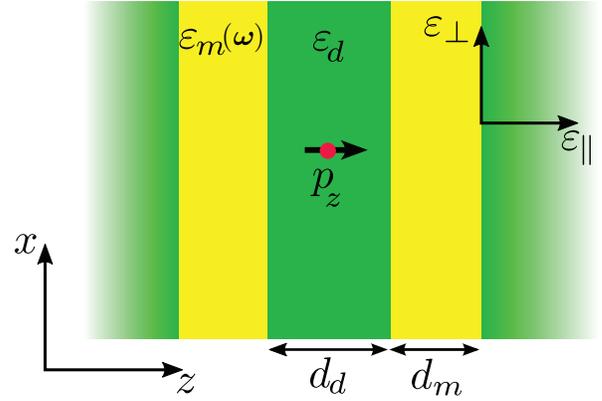}
\caption{(Color online) Geometry of the structure: hyperbolic metamaterial formed by metal and dielectric layers with permittivities $\varepsilon_d, \varepsilon_m$ and thicknesses $d_d, d_m$, correspondingly. The dipole is placed inside the layered structure, $z$-component of dipole moment is oriented parallel to anisotropy axis.}
\label{pic:system}
\end{center}
\end{figure}

We consider a  layered structure, consisting of alternating layers of metal and dielectric with dipole placed in the centre of one of the dielectric layers (see Fig.~\ref{pic:system}). Such structure can be described as an effective medium with following dielectric constants:
\begin{equation}
\varepsilon_{\perp} = \frac{d_m \varepsilon_m + d_d \varepsilon_d}{d_m + d_d}, ~ \frac{1}{\varepsilon_{\parallel}} = \frac{1}{d_m + d_d} \left(\frac{d_m}{\varepsilon_m} + \frac{d_d}{\varepsilon_d} \right) ,
\end{equation}
where $d_d$, $d_m$ are thicknesses and $\varepsilon_d$, $\varepsilon_m$ are permittivities of layers; indexes $\perp$ and $\parallel$ indicates components perpendicular and parallel to the anisotropy axis, respectively (so in Cartesian coordinates $\varepsilon_{xx} = \varepsilon_{yy} \equiv \varepsilon_{\perp}$ and $\varepsilon_{zz} \equiv \varepsilon_{\parallel}$). Permittivity of metal can be described within the Drude model:
\begin{equation}
\varepsilon_m = \varepsilon_{\infty} - \frac{\omega_p^2}{\omega^2 + \omega \gamma i}
\end{equation}

To achieve hyperbolic medium regime the condition $\mathrm{Re}\varepsilon_{\perp} \mathrm{Re}\varepsilon_{\parallel} < 0$ should be satisfied. Depending on the sign of $\varepsilon_{\perp}$ and $\varepsilon_{\parallel}$ hyperbolic metamaterials could be separated on two types: Type ~I with $\varepsilon_{\perp} > 0, \varepsilon_{\parallel} < 0$ and Type~II with $\varepsilon_{\perp} < 0, \varepsilon_{\parallel} > 0$. If both of dielectric constants are positive the isofrequency contour is ellipse as in usual anisotropic materials.

The eigenfrequency of the dipole $\omega$ can be calculated using following expression \citep{poddubny2011spontaneous}:
\begin{equation}
\omega - \omega_0 = \frac{4 \pi q_0^2 d^2}{\hbar} \int \frac{d^3 k}{(2 \pi)^3} ~~ G_{\textbf{\textit{k}}, zz} ~ e^{-k^2 a^2},
\label{eq:podd}
\end{equation}
where $\omega_0$ is the resonance frequency, $q_0 = \omega/c$, $a$ is the dipole characteristic size,  $d=\xi e a$ is the effective matrix element of the dipole moment, where $\xi\in[0,1]$ is the geometric factor. The Gaussian factor stands for the spatial extent of the dipole, $\mathbf{k}$ is a wave-vector and $G_{\textbf{\textit{k}}, zz}$ is the Green's function (we consider dipole orientation along $z$ axis):
\begin{equation}
G_{\textbf{\textit{k}}, zz} = \frac{1}{\varepsilon_{\parallel}} \frac{1 - k_{\|}^2/(q_0^2 \varepsilon_{\perp})}{q_0^2 - k_{\perp}^2/\varepsilon_{\parallel} - k_{\parallel}^2/\varepsilon_{\perp}}.
\end{equation}
The right-hand side of equation \eqref{eq:podd} can be considered as function $R$, which in general depends on $\omega$, but conventionally $R(\omega)$ is assumed to be equal to $R(\omega_0)$ which corresponds to the weak-coupling regime. However, hyperbolic metamaterial is highly dispersive material and we suppose this dispersion should be included in calculation to get the right result. To prove this statement we rewrite the equation \eqref{eq:podd} for case $R(\omega) \neq R(\omega_0)$ so it becomes 
\begin{equation}
x - 1 =  
\frac{\alpha \tilde{a}^2 \xi^2}{\pi\varepsilon_{\parallel}} \int dkd\theta~\tilde{k}^2 \sin\theta  \frac{x^2 - \tilde{k}^2 \cos^2 \theta / \varepsilon_{\perp}}{x^2 - \tilde{k}^2 / \chi(\theta)}e^{-\tilde{k}^2\tilde{a}^2},
\label{eq:num}
\end{equation}
where complex value  $x = \omega/\omega_0$, $\tilde{a} = \omega_0 a/c$, $\tilde{k} = k/(\omega_0/c)$, $\alpha=e^2/(\hbar c)\approx 1/137$ is the fine structure constant,  $\theta$ is the angle between $z$-axis and wave-vector $\mathbf{k}$ (we consider spherical coordinates) and 
\begin{equation}
\chi(\theta) = \left(\frac{\sin^2\theta}{\varepsilon_{\parallel}} + \frac{\cos^2\theta}{\varepsilon_{\perp}} \right)^{-1}.
\end{equation}
The permittivity of metal layers now is the function of $x$:
\begin{equation}
\varepsilon_m = \varepsilon_{\infty} - \frac{\omega_p^2}{(x \omega_0)^2 + x \omega_0 \gamma i},
\end{equation}
and therefore $\varepsilon_{\perp}$, $\varepsilon_{\parallel}$ are also functions of $x$.

Purcell factor can be calculated as imaginary part of relation $x/x_{vac}$, i.e. $F_p = \text{Im}[x/x_{vac}]$, and Lamb shift as $L_s = \text{Re}[x] - 1$, where $x_{vac}$ is the solution of \eqref{eq:num} in the case of dipole placed in vacuum:
\begin{equation}
x_{vac} - 1 = -\frac{ \alpha \tilde{a}^2 \xi^2}{6} \left(\frac{4 \tilde{a}^2 - 1}{\sqrt{\pi} \tilde{a}^3} + 4 e^{-\tilde{a}^2} \left[i - \text{erfi}(\tilde{a})\right] \right).
\end{equation} 
The numerical solution of Eq.~\eqref{eq:num} can be performed iteratively. At the first iteration we solve the equation setting $x_0=1$ in the right hand-side of the equation. This gives us the weak-coupling result $x_1$ obtained in~\cite{poddubny2011spontaneous}. At the next iteration we substitute the solution for $x$ to the right-hand side of the equation and obtain the second order correction. The procedure is repeated until the convergence condition is fulfilled $|x_{n}-x_{n-1}|/|x_{n-1}|<\delta$, where $\delta$ is the convergence threshold parameter which was set to $10^{-8}$ in the calculation.
To obtain the analytical approximations for the corrections to the Purcell factor it is instructive to consider the region in the vicinity of the two topological transitions $\varepsilon_{\perp}\approx 0$ and $\varepsilon_{\|}\approx 0$. For the case of first topological transition we first assume that the dipole bare frequency is tuned exactly to the frequency of the topological transition $\varepsilon_{\perp}(\omega_0)=0$. We also neglect contribution of the imaginary part of the permittivities. The expansion of the permittivity tensor in the vicinity of the topological transition is given by:
\begin{equation}
\begin{aligned}
&\varepsilon_{\perp}=(\varepsilon_{\infty}+\varepsilon_d)(x-1),\\
&\varepsilon_{\|}=-\varepsilon_d^2\left[(x-1)(\varepsilon_{\infty}+\varepsilon_d)\right]^{-1}
\end{aligned}
\end{equation}
We then apply the first order correction. The right hand side of the equation~\eqref{eq:num} in this case is exactly equal to zero leading to the trivial solution $x=1$ which coincides with the zero-order approximation. Thus, in the case of the absence of losses at the low frequency topological transition $\varepsilon_{\perp}=0$ the self-consistent approach will fully coincide with the conventional approach. It should be taken into account though that introducing the losses to the system will result in the non-zero right-hand side of the Eq.~\eqref{eq:num} leading to the discrepancy between the two approaches as will be seen in the numerical calculations.

The second topological transition takes place at $\varepsilon_{\|}\approx 0$. In the vicinity of this point the permittivities can be expanded with respect to $(x-1)$ as:
\begin{equation}
\begin{aligned}
&\varepsilon_{\perp}=\varepsilon_{d}/2,\\
&\varepsilon_{\|}=4\varepsilon_{\infty}(x-1).
\end{aligned}
\end{equation}
Contrary to this case, the right hand side of Eq.~\eqref{eq:num} diverges logarithmically at $x\approx 1$. Namely it can be written to the lowest order in $x-1$ as

\begin{align}
\xi\alpha\tilde{a}^2\left(\frac{1+\frac{1}{2}\ln\left(\frac{2\varepsilon_{\infty}(x-1)}{\varepsilon_d}\right)}{\varepsilon_d\tilde{a}^3\sqrt{\pi}}+\frac{1}{2}i\sqrt{2\varepsilon_d}\right).
\end{align}
This should lead to the large discrepancies between the self-consistent and the conventional approach even in the presence of losses which will be demonstrated in numerical calculations.

\section{RESULTS}
\begin{figure*}
 \begin{center}
  \includegraphics[width=0.9\linewidth]{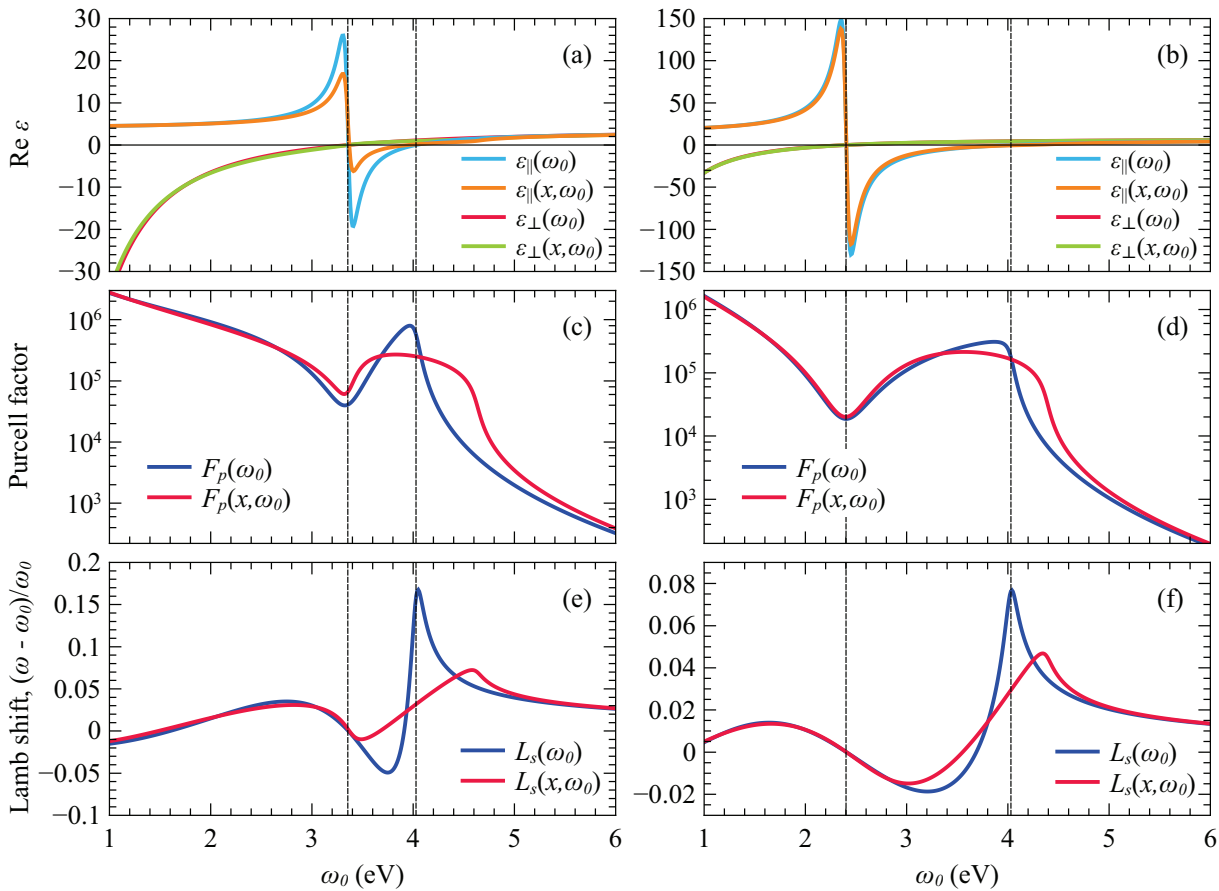}
  \caption{(Color online) Spectrum of (a,b) effective parameters, (c,d) Purcell factor and (e,f) Lamb shift. Left column (a,c,e) corresponds to $\varepsilon_d = 2.2$ and the right column (b,d,f) to $\varepsilon_d = 9$. Other parameters of the structure are given in the text. Values labeled as functions of $\omega_0$ are obtained within the first order expansion approach, in contrast to the values labeled as functions of $x$ and $\omega_0$, which are obtained within the self-consistent approach. Vertical lines, which pass through the points $\varepsilon_{\perp}(\omega_0) = 0$ and $\varepsilon_{\parallel}(\omega_0) = 0$, indicate topological transitions.}
  \label{pic:results}
 \end{center}
\end{figure*}

Results of calculations are shown on Fig.~\ref{pic:results}. Parameters of the system are following: $d_m~=~d_d~=~20$~nm, $\varepsilon_{\infty}~=~4.96$, $\omega_p~=~8.98$~eV, $\gamma~=~0.1$~eV. Left column corresponds to  $\varepsilon_d=2.2$ and right to  $\varepsilon_d=9$.  Plots of the dielectric permittivities components are shown in Figs.~\ref{pic:results}(a,b).  We are interested in small dipoles, like quantum dots or dye molecules, so the size of the dipole was chosen to be $a~=~2.2$~nm. We note that in Figs.~\ref{pic:results}(a,b) we plot the effective parameters spectrum both at bare dipole frequencies and at the frequencies renormalized due to the interaction with the back-reflected electromagnetic fields. These plots are labeled as $\varepsilon_{\sigma}(x,\omega_0)$, $\sigma=\|,\perp$ in contrast to the bare frequency plots labeled as $\varepsilon_{\sigma}(\omega_0)$. It can be seen that the two spectra types show the discrepancies only in the hyperbolic regime and especially in the vicinity of the  hyperbolic transition $\varepsilon_{\|}=0$ as was discussed in the previous section.
This can be also seen in the Purcell factor spectra shown in Figs.~\ref{pic:results}(c,d).
A sharp dip in Purcell factor corresponds to the transition to $\varepsilon_{\perp}=0$ regime and signifies the topological transition from the single-sheet to double-sheet hyperbolic regime. It can be also observed that the sharp difference between the first order calculation and the self-consistent approach occurs in the double sheet hyperbolic region $\varepsilon_{\perp}>0,\varepsilon_{\|}<0$ and maximizes at the topological transition $\varepsilon_{\|}\approx 0$. The same picture can be observed in the Lamb shift spectra - the most pronounced differences are observed in the vicinity of the topological transition $\varepsilon_{\|}\approx 0$. 

\begin{figure}[ht!]
\begin{center}
\includegraphics[width=0.9\linewidth]{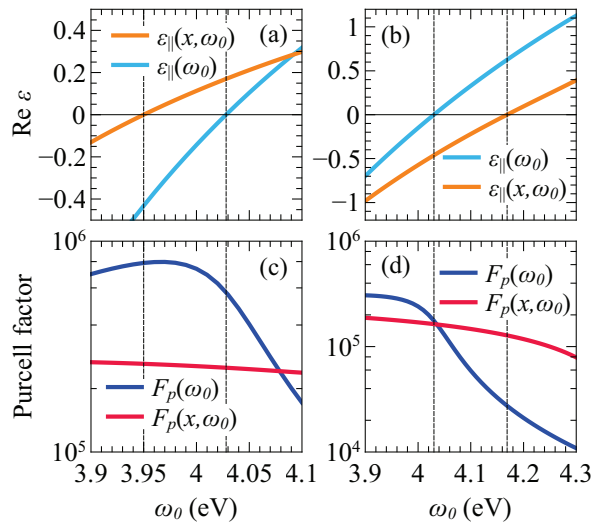}
\caption{(Color online) Spontaneous topological transition (a,b) between hyperbolic and elliptical regimes (vertical lines pass through the points where $\varepsilon_{\parallel} \approx 0$). Values of Purcell factor (c,d) and effective parameters at bare dipole frequency are labeled as $\varepsilon_{\parallel}(\omega_0)$ and $F_p(\omega_0)$, respectively; $\varepsilon_{\parallel}(x,\omega_0)$ and $F_p(x,\omega_0)$ correspond to values obtained for the frequencies renormalized due to the interaction with the back-reflected electromagnetic fields. Permittivities of dielectric layers are (a,c) $\varepsilon_d = 2.2$  and (b,d) $\varepsilon_d = 9$; other parameters and details are given in the text.}
\label{pic:transition}
\end{center}
\end{figure}

In order to understand the nature of these discrepancies it is instructive to plot the effective material parameters of the metamaterial, i.e. dielectric permittivities at the effective emitter frequencies accounting for the Lamb shift, and not the bare frequency. These are plots for two values of dielectric permittivity $\varepsilon_d$ are shown in Figs.~\ref{pic:transition}(a-d). We can see that there are two distinct scenarios in here. In the case of the small dielectric permittivity $\varepsilon_d=2.2$ even when the dipole is effectively in the hyperbolic regime (the region between two dotted lines at Fig.~\ref{pic:transition}(a) the renormalization of the dipole frequency leads to the effective positive $varepsilon_{\|}$, and thus dipole effectively radiates in the elliptic regime which sufficiently decreases its Purcell factor as can be seen at Fig.~\ref{pic:transition}(c). Contrary to this case, for the large permittivity $\varepsilon_d=9.0$ the trend is opposite and dipole effectively radiates in the hyperbolic regime even when the bare frequency is in the elliptic regime. We should also note that the described  discrepancies increase further with decreasing of the effective emitter size. 
It should be noted that the effect observed is the example of the spontaneous topological transition in the sense, that unlike previously reported topological transitions in hyperbolic metamaterials~\cite{krishnamoorthy2012topological} here the topological transition occurs not due to the external change of the parameters (emitter frequency, temperature or any other) but only due to the interaction of the emitter with vacuum electromagnetic field fluctuation (i.e. due to the Lamb shift). These effects of course could not be traced in the conventional first-order expansion approach however as we have shown could play a significant role in the emitter radiation in the highly dispersive metamaterials.

\section{CONCLUSION}
We have shown that in the highly dispersive metamaterial, i.e. when the effective parameters depend sharply on frequency, the conventional approach for calculation of the Purcell effect and Lamb shift lead to the incorrect results and a rigorous self-consistent approach accounting for the effective dipole frequency shift should be used instead. We have shown that the largest discrepancies arise at the vicinity of the topological transition from elliptical to hyperbolic regime and predicted the effect of the Lamb shift induced spontaneous topological transition. These findings are essential for the  robust design of the optoelectronic devices based on metal-dielectric hyperbolic metamaterials.

\section{Acknowledgements}
We thank A.N. Poddubny for helpful discussions. I.I.   appreciates the support
of the Ministry of Education and Science of the
Russian Federation (Zadanie No. 3.1231.2014/K), Grant
of the President of Russian Federation (MK-5220.2015.2)
and RFBR project 16-32-60123

\bibliography{purcell_draft_v3}

\end{document}